\documentclass[aps,prl,twocolumn,showpacs,superscriptaddress,amsmath,amssymb]{revtex4-1}
\usepackage{graphicx}
\usepackage{dcolumn}
\usepackage{bm}
\usepackage{amsmath}
\usepackage{amssymb}
\usepackage{graphicx}
\usepackage{indentfirst}
\usepackage{booktabs}
\usepackage{multirow}
\usepackage{colortbl}
\usepackage{epsfig}

\selectfont

\begin{document}
\title{Intrinsic anomalous spin Hall effect}

\author{Ping Li}
\address{State Key Laboratory for Mechanical Behavior of Materials, Center for Spintronics and Quantum System, School of Materials Science and Engineering, Xi'an Jiaotong University, Xi'an, Shaanxi, 710049, China}
\address{Key Laboratory for Computational Physical Sciences (Ministry of Education), Fudan University, Shanghai, 200433, China}
\author{Jing-Zhao Zhang}
\address{State Key Laboratory for Mechanical Behavior of Materials, Center for Spintronics and Quantum System, School of Materials Science and Engineering, Xi'an Jiaotong University, Xi'an, Shaanxi, 710049, China}
\author{Zhi-Xin Guo}
\email{zxguo08@xjtu.edu.cn}
\address{State Key Laboratory for Mechanical Behavior of Materials, Center for Spintronics and Quantum System, School of Materials Science and Engineering, Xi'an Jiaotong University, Xi'an, Shaanxi, 710049, China}
\address{Key Laboratory for Computational Physical Sciences (Ministry of Education), Fudan University, Shanghai, 200433, China}
\author{Tai Min}
\address{State Key Laboratory for Mechanical Behavior of Materials, Center for Spintronics and Quantum System, School of Materials Science and Engineering, Xi'an Jiaotong University, Xi'an, Shaanxi, 710049, China}
\author{X. R. Wang}
\email{phxwan@ust.hk}
\address{Physics Department, The Hong Kong University of Science and Technology, Clear Water Bay, Kowloon, Hong Kong.}
\address{HKUST Shenzhen Research Institute, Shenzhen 518057, China.}

\date{\today}
\begin{abstract}
Charge-spin interconversion in magnetic materials is investigated by
using first-principles calculations. In addition to the conventional spin
Hall effect (SHE) that requires mutual orthogonality of the charge current,
spin-flow direction, and spin polarization, the recently proposed
anomalous SHE (ASHE) is confirmed in Mn$_2$Au and WTe$_2$. The interaction
of the order parameter with conduction electrons leads to sizeable nonzero
spin Berry curvatures that give rise to nonzero anomalous spin Hall
conductivity (ASHC). Our calculations show that the ASHE is intrinsic and
originates from the order-parameter-controlled spin-orbit
interaction, which generates an extra anomalous effective field.
A useful relationship among the order parameter, spin Berry curvature and
ASHC is revealed. Our findings open a new avenue for arbitrary-type
spin current generation and detection. 
\end{abstract}

\maketitle
\section*{INTRODUCTION}
The spin Hall effect (SHE) is related to spin current generation or spin
accumulation from an electric current\cite{Zhang,MacDonald,
Jungwirth,Hirsch}. Different from an electric current, which
is a flow of charges and a vector, a spin current, or an angular momentum flow, is
a tensor of rank two with nine components in Cartesian coordinates.
The SHE has attracted great attention and is widely used to generate
a spin current, which is an indispensable component of spintronics
\cite{Zhang,MacDonald,Jungwirth,Hirsch}. The inverse SHE is a
powerful method for detecting spin currents \cite{Chien,Deranlot}.

In the SHE, a charge current along the $\hat \beta$ direction in a material
with a strong spin-orbit interaction (SOI) can generate a spin current
$j^{\gamma}_{\alpha}$ whose propagation and spin polarization are
along the $\hat \alpha$ and $\hat \gamma$ directions, respectively.
The conventional SHE requires $\hat \alpha$, $\hat \beta$, and $\hat
\gamma$ to be mutually orthogonal to each other. Recently, Wang proposed
that interactions of conduction electrons with order parameters, such
as magnetization for ferromagnets and the N$\acute{e}$el order for
antiferromagnets, give rise to the anomalous SHE (ASHE) \cite{Wang}.
In linear response theory, spin current $j^j_i$ due
to charge current $\bf J$ is given by
\begin{equation}
j^j_i= \frac{\hbar}{2e} \theta ^{SH}_{ijk} J_k,
\end{equation}
where $\theta ^{SH}_{ijk}$ is the spin Hall angle tensor of rank 3.
$i, j, k=1,2,3$ denote the x, y, and z directions, and the Einstein
summation notation is used. $\emph{i}$, $\emph{j}$, and $\emph{k}$
denote the spin current direction, spin polarization direction, and
charge current direction. Based on the general tensor requirement
of a physical quantity, one has \cite{Wang}
\begin{small}
\begin{equation}
\begin{split}
\theta ^{SH}_{ijk} =\theta _{0} \epsilon _{ijk} + [(\theta _{1} +
\theta _{2})\delta _{ij} \delta_{kl \neq i} + \theta _{1}\delta _{ik}
\delta_{jl \neq i} + \theta _{2} \delta _{il} \delta_{jk \neq i}]M_l,
\end{split}
\end{equation}
\end{small}
where \emph{M$_l$} and $\epsilon _{ijk}$ are the $\emph{l}$th component
of $\textbf{M}$ and the Levi-Civita symbols, respectively.
$\theta _{0}$ is the usual spin Hall angle that does not depend
on $\textbf{M}$, while $\theta _{1}$ and $\theta _{2}$ are two
ASHE coefficients. Thus, the conventional SHE is nonzero only
when $i, j,$ and $k$ take different values. However, the terms involving
\emph{M$_l$} can be nonzero, which leads to nonzero ASHE
spin currents, as indicated in Figs. 1f-1h. This newly proposed
ASHE was not only directly verified by experiments of the anomalous
spin torques generated in Mn$_{2}$Au \cite{Pan} but also agreed
with the details of the observed magnetization-dependent anomalous
inverse SHE (AISHE) in a Pt/Co/Pt heterostructure \cite{Chuang},
TbCo alloys \cite{Yagmur}, and Co/Pd multilayers
\cite{Dinghai}. The ASHE overcomes the limitation of the SHE, and
the generated spin current can be controlled by order parameters.
In particular, the ASHE allows the spin polarization and spin
propagation directions to be collinear when the applied electric
current is along the order parameter direction. Such a spin current
can generate an out-of-plane anti-damping spin-orbit torque (SOT)
that is crucial to realizing field-free switching of perpendicular
magnetization in high-density SOT devices such as SOT-MRAM
\cite{Gaudin,Buhrman,Yang,Ralph,yin,Chen}.

Although the ASHE concept has been widely recognized, its microscopic
origin remains to be revealed. Recently, first-principles calculations
\cite{Haney,Felser} were used to study the anomalous spin torques observed
in experiments involving ferromagnets and antiferromagnets. However, the
discussions are limited to qualitative symmetry analysis \cite{Ebert},
and the intrinsic mechanism of the ASHE is still elusive \cite{Park}.
One cannot quantitatively compute the anomalous spin torques of specific
materials without knowing the intrinsic mechanism. Such an investigation
is particularly desirable in the field of high-performance SOT devices
\cite{Ralph,Chen,Park}.

Using the local spin density approximation theory with the SOI, we
investigate the intrinsic mechanism of the ASHE, with the following findings.
1) The ASHE originates from the order-parameter-induced extra
spin-dependent electric field, which generates anomalous spin currents via the SOI.
2) The anomalous spin Hall conductivity (ASHC), which quantitatively
represents the strength of the ASHE, can be well described by the spin
Berry curvature. 3) There is an intrinsic connection among the order parameter,
spin Berry curvature, and ASHC.

\section*{METHODS}
Our spin-polarized density functional theory (DFT) calculations are performed using the Vienna
{\it ab initio} Simulation Package (VASP) \cite{26,27} and QUANTUM
ESPRESSO (QE) \cite{QE1,QE2}, in which the projector augmented wave
method and a plane-wave basis set are applied. We use the VASP
for Mn$_2$Au and WTe$_2$ and QE for Fe. All calculations are
performed using spin polarization within the framework of the local
spin density approximation and generalized gradient approximation
for the exchange-correlation energy. The electron exchange-correlation
functional is described by the Perdew-Burke-Ernzerhof functional of the generalized gradient approximation
\cite{28}. The plane-wave
cutoff energy is chosen to be 500 eV for Mn$_2$Au and WTe$_2$ and
120 Ry for Fe. Moreover, $\Gamma$-centered $k$ meshes of $20\times
20\times 20$ (Pt), $16\times 16\times 16$ (Fe), $18\times 9\times 3$
(WTe$_2$), and $24\times 24\times 8$ (Mn$_2$Au) are adopted in the
self-consistent calculations. To better describe the van der Waals
(vdW) interaction, the optB86b-vdW functional is adopted for WTe$_2$
\cite{29,30}. Then, DFT wave functions are projected onto maximally
localized Wannier functions using the WANNIER90 package \cite{31},
and the Kubo formula is applied to calculate the spin Hall conductivity (SHC) \cite{Guo,Zhao}.
Dense k-point meshes of $150\times 150\times 150$, $600\times 600
\times 600$, $200\times 200\times 200$ and $150\times 150\times 150$
are employed for Pt, Fe, WTe$_2$ and Mn$_2$Au, respectively, to
perform the Brillouin zone (BZ) integration for the SHC calculations with a careful
convergence test.

\section*{RESULTS}

According to local spin density approximation theory \cite{LSDA1,LSDA2},
a position-dependent spin-dependent electric field $\vec E^p (\vec r)$,
which is proportional to the difference in the charge density gradients of
spin-up and spin-down electrons, [$\nabla n^{\uparrow}(\vec r)$] and
[$\nabla n^{\downarrow}(\vec r)$], can be induced by the magnetic order
parameter $\bf M$ (see Supporting Information).
Note that the magnetization-induced spin-dependent
electric field differs from the conventional electric field.
The conventional electric field is defined as $\vec E(\vec r)=-[\nabla
H^{\uparrow}(\vec r)+\nabla H^{\downarrow}(\vec r)]$, while the
spin-dependent electric field is defined as
$\vec E^p (\vec r)= -[\nabla H^{\uparrow}(\vec r)-\nabla
H^{\downarrow}(\vec r)]$, where $H^{\uparrow}$ and $H^{\downarrow}$
are the Hamiltonians for spin-up and spin-down electrons, respectively.
Hence, the spin-dependent electric field represents the difference in the
electric fields between spin-up and spin-down electrons.

In a collinear magnetic structure with inversion symmetry, the
effective spin-dependent electric field (averaged over the unit cell) defined
as $\vec E_{eff}^{p}=\int \vec E^p (\vec r) d \vec r =0$ is zero.
However, $\vec E_{eff}^{p}$ becomes nonzero if the inversion
symmetry is broken. Such an $\vec E_{eff}^{p}$ can induce an effective
Hamiltonian $H^{eff}_{SOC}$ via the SOI (see Supporting Information),
\begin{equation}
H^{eff}_{SOC} =\frac{\hbar e}{2mc^2} \left (\vec {\sigma} \times
\vec E_{eff}^{p} \right ) \cdot \vec v,
\end{equation}
where $\vec \sigma$ and $\vec v$ are the Pauli matrices and electron
velocity, respectively. The nonzero  $H^{eff}_{SOC}$
means that the spin-up and spin-down electrons experience different
forces and thus have different velocities.
Hence, Eq. (3) leads directly to the ASHE as long as
$\vec \sigma$ is perpendicular to $\vec E_{eff}^{p}$.
For example, in a magnetic material without inversion symmetry in
the $\vec z$ direction, i.e., $E_{eff}^{p_z}\ne 0$, one has nonzero
$H^{eff}_{SOC} =\frac{\hbar e}{2mc^2} E_{eff}^{p_z}\left ( {\sigma_y}
v_x-\sigma_x v_y \right )$. Thus, energy splitting between spin-up
and spin-down electrons is generated by the SOI, leading to different
velocities. As a result, anomalous spin currents with spin polarization
perpendicular to $\vec z$, as predicted in Wang's theory \cite{Wang},
will be generated (see Supporting Information). Note that
since the spin-dependent electric field directly correlates to the
magnetic order parameter $\bf M$, nonzero $E_{eff}^{p}$ is naturally
expected in the direction of broken inversion symmetry induced
by $\bf {M}$, rather than in the direction of the magnetic moment.

We find that the strength of the ASHE can be quantitatively described
by the spin Berry curvature. According to the Kubo formula,
the SHC $\sigma^\gamma_{\alpha \beta}$ of
spin polarization direction $\hat \gamma$ and spin flow direction
$\hat \alpha$ under an external electric field (charge current)
along $\hat \beta$ can be expressed as \cite{Guo,Zhao,Luqi,Leandro}
\begin{equation}
\sigma^\gamma_{\alpha \beta}= -\frac{e^2}{\hbar} \frac{1}{VN^3_k}
\sum_{k}\sum_{n}f_{nk}\Omega^\gamma_{n,\alpha \beta}(\bf{k}),
\end{equation}
where V is the cell volume and $N^3_k$ is the number of $k$ points
in the BZ. $f_{nk}$ represents the Fermi-Dirac
distribution function, and $\Omega^\gamma_{n,\alpha \beta}(\emph{\textbf{k}})$ is the band-projected spin Berry curvature,
defined as
\begin{widetext}
\begin{equation}
\Omega^\gamma_{n,\alpha \beta}({\bf k})=\hbar^2\sum_{m \neq n}\frac{-2\mathrm{Im}[\left \langle n{\bf k} \mid \frac{1}{2} \{\hat
\sigma_\gamma,\hat \upsilon_\alpha \} \mid m{\bf k} \right \rangle
\left \langle m {\bf k} \mid \hat \upsilon_\beta \mid n{\bf k}
\right \rangle]}{(\epsilon_{n\bf k}-\epsilon_{m\bf k})^2}.
\end{equation}
\end{widetext}
From Eqs. (4) and (5), one can see that the Hall conductivity
$\sigma^\gamma_{\alpha \beta}$ is intrinsically determined by
the band structure through the spin Berry curvature
$\Omega^\gamma_{n,\alpha \beta}(\bf{k})$ and density of states.
In the conventional SHE, nonzero $\Omega^\gamma_{n,\alpha \beta}(\emph{\textbf{k}})$ is expected only when $\alpha$, $\beta$ and
$\gamma$ are mutually orthogonal to each other. However, as
discussed below for the case involving the order parameter,
nonzero $\Omega^\gamma_{n,\alpha \beta}(\emph{\textbf{k}})$ and
thus a sizeable ASHE can also be obtained when only two of them
are mutually perpendicular. In addition, the general
Berry curvature formula, which requires $\alpha \neq \beta$, was
derived by simply considering the spin direction perpendicular
to the velocities of the spin current and charge current
($\alpha \neq \beta \neq \gamma$) \cite{Niu}. In this case, the
spin degree of freedom is not effectively considered, and the
Berry curvature is in fact a tensor of rank 2. However, in the
definition of spin Berry curvature shown in Eq. (5), the
interaction between the spin degree of freedom and spin current
is effectively considered via the anticommutator operator.
As a result, the spin Berry curvature becomes a tensor of rank 3,
which is nonzero even for $\alpha = \beta$.

\begin{figure}[htb]
\begin{center}
\includegraphics[angle=0,width=0.95\linewidth]{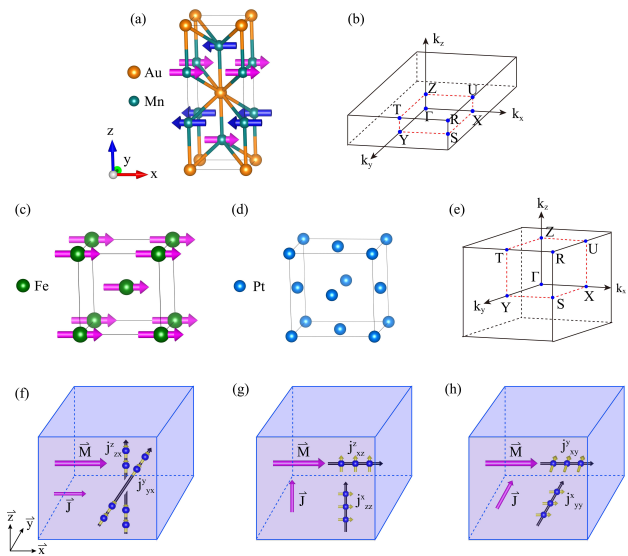}
\caption{ \textbf{Atomic structures, schematic diagrams of the BZs,
and schematic diagrams of the ASHE.} (a), (c) and (d) show the atomic
structures of TET Mn$_{2}$Au, BCC Fe and FCC Pt, respectively.
(b) and (e) show the corresponding BZs. Note that BCC Fe and FCC Pt
have the same BZ structure, shown in (e). (f) illustrates the ASHE
with charge current $\vec J$ parallel to order parameter $\bf M$,
and (g, h) present the ASHE with charge current $\vec J$ perpendicular
to $\bf M$.}
\end{center}
\end{figure}

We first consider tetragonal (TET) Mn$_{2}$Au as an illustration, whose nonzero ASHE has been experimentally observed
\cite{Pan}. We systematically show below an intrinsic
connection among the order parameter, spin Berry curvature and ASHE.
DFT calculations were performed to obtain
the spin Berry curvature and SHC (see Supporting Information). The crystal structure and
schematic BZ diagram of Mn$_{2}$Au are illustrated in Figs. 1a and 1b,
which verify that Mn$_{2}$Au is an antiferromagnetic material with a
N$\acute{e}$el order parameter $\bf M$. In this study, $\bf M$ is
fixed in the $\vec x$ direction. Hence, as indicated in Figs. 1e-1g,
six nonzero ASHCs are expected, that is, $\sigma^y_{yx}$ and
$\sigma^z_{zx}$ for charge current $\bf J$ parallel to $\bf M$ and
$\sigma^x_{zz}$, $\sigma^z_{xz}$, $\sigma^x_{yy}$, and $\sigma^y_{xy}$
for $\bf J$ perpendicular to $\bf M$ \cite{Wang}.

The calculated components of both the conventional SHC
(CSHC) and ASHC are shown in Table I.
In addition to the large CSHCs, there are also sizeable ASHCs in
Mn$_{2}$Au. Moreover, both CSHCs and ASHCs exhibit significant
anisotropy, depending on the combination of the spin current, spin
polarization, and charge current directions. This characteristic
is quite different from that of the conventional SHE, whose
anisotropy arises solely from the crystalline structure \cite{Ebert}.
For instance, the symmetry of Mn$_{2}$Au is expected to give rise
to the same CSHCs for $\sigma^{z}_{xy}$ and $\sigma^{z}_{yx}$.
However, the former [270.58 $(\hbar/e)S/cm$] is more than 3 times
larger than the latter [-81.11 $(\hbar/e)S/cm$] because of the
broken fourfold rotational symmetry due to $\bf M$ along
the $\vec x$ direction.

Moreover, $\bf M$ along the $\vec x$ direction in Mn$_{2}$Au can lead to an
additional nonzero spin-dependent electric field in the $\vec z$
direction ($E^{p}_z$) due to its broken inversion symmetry. According to Eq. (3), the nonzero $E^{p}_z$ would induce a
spin current with spin polarization in both the $\vec x$ and
$\vec y$ directions via the SOI and thus leads to sizeable ASHCs.
The details are given in the Supporting Information.
Our DFT calculations indeed show that the absolute values of ASHCs
with spin polarization in the $\vec x$ direction ($\sigma^x_{zz}$,
$\sigma^x_{yy}$) and $\vec y$ direction ($\sigma^y_{yx}$, $\sigma^y_{xy}$)
are significantly larger than those with spin polarization in the
$\vec z$ direction ($\sigma^z_{zx}$, $\sigma^z_{xz}$) (Table I).
The nonzero values of $\sigma^z_{zx}$ and $\sigma^z_{xz}$ may be
from other ASHE mechanisms that are beyond the scope of effective
field theory discussed above. Among the six ASHC components,
the largest component is $\sigma^{y}_{yx}$ [15.57 $(\hbar/e)S/cm$],
the magnitude of which reaches 19\% of that of its conventional
counterpart $\sigma^{z}_{yx}$ [-81.11 $(\hbar/e)S/cm$].
Thus, a sizeable spin current, whose propagation direction is
collinear to the spin polarization direction, can be effectively
generated in Mn$_{2}$Au. This result is consistent with the
experiments \cite{Pan}. The good agreement between our
theoretical derivations and DFT calculations verifies the intrinsic
connection among the order parameter, spin Berry curvature and ASHE.

\begin{table}[htbp]
\caption{ \textbf{Spin Hall Conductivity.} Calculated CSHCs and ASHCs
for Fe, Mn$_2$Au and Pt. Each SHC is denoted as $\sigma^{\gamma}_{
\alpha\beta}$, where $\alpha$ is the direction of the spin current,
$\beta$ is that of the applied electrical field and $\gamma$ is that
of the spin polarization of the spin current (unit: $(\hbar/e)S/cm$).
The magnetic order parameter $\bf M$ is fixed to be along the
$\vec x$ direction for Fe and Mn$_2$Au.}
\begin{tabular}{cccccccc}
  \hline
  & SHC        & Mn$_2$Au     & Fe        & Pt              \\
  \hline
 & $\sigma^{z}_{xy}$   & 270.58     & 120.76    & 2239.42     \\
     \raisebox{1ex}{CSHC}
& $\sigma^{z}_{yx}$    & -81.11       & -683.71   & -2239.58   \\
  \hline
 & $\sigma^{y}_{yx}$    & 15.57        &  2.87     & 0.00    \\
 & $\sigma^{z}_{zx}$     & 0.02      & -2.89     & 0.00     \\
& $\sigma^{x}_{zz}$    & -10.46       & -0.66     & 0.00     \\
     \raisebox{1ex}{ASHC}
 & $\sigma^{z}_{xz}$    & -2.95    & -3.14     & 0.00     \\
& $\sigma^{x}_{yy}$    & -6.72        &  0.49     & 0.00    \\
  & $\sigma^{y}_{xy}$     & -9.09        &  1.37     & 0.00  \\
  \hline
\end{tabular}
\end{table}

As a comparison, we also calculated the SHCs
of bulk Fe in the body-centered cubic (BCC) structure.
As shown in Fig. 1(c), although a large $\bf M$ exists along the
$\vec x$ direction, $\vec E^{p}$ in this system is zero due to
the inversion symmetry (see Supporting Information). This means that the order
parameter does not directly induce the ASHE in BCC Fe.
Our DFT calculations show that despite its considerably large CSHCs,
all the ASHCs in Fe are very small, with an ASHC/CSHC ratio of
less than 0.5\% (Table I). We also explored the
SHCs of nonmagnetic Pt in the face-centered
cubic (FCC) structure. As shown in Table I, although the CSHCs of
Pt are larger than those of Mn$_{2}$Au by an order of magnitude,
all its ASHCs are negligible [$\textless 10^{-3}$ $(\hbar/e)S/cm$].
This result confirms the intrinsic connection between the order
parameter and ASHE.

\begin{figure}[htb]
\begin{center}
\includegraphics[angle=0,width=0.70\linewidth]{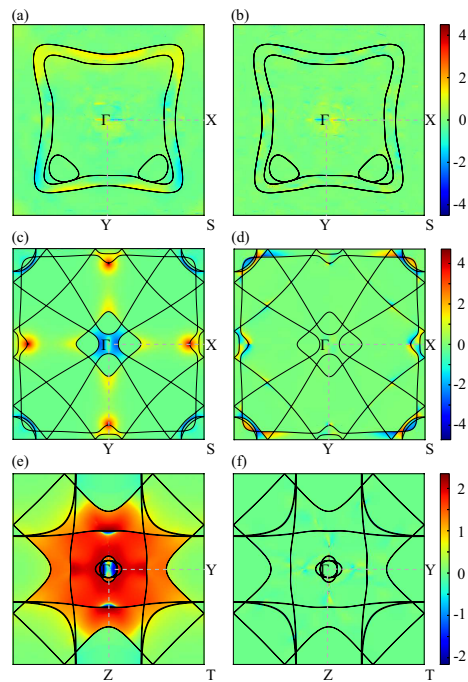}
\caption{\textbf{Spin Berry curvature in BZ slices.} Density plots
of the spin Berry curvature in the BZ slice of k$_z$ = 0 for bulk Mn$_{2}$Au
(a, b) and Fe (c, d) and the 2D BZ of k$_x$ = 0 for Pt (e, f).
(a), (c), and (e) are the spin Berry curvature for CSHCs: $\sigma^z_{yx}$ of
Mn$_{2}$Au, $\sigma^z_{xy}$ of Fe, and $\sigma^z_{xy}$ of Pt,
respectively. (b), (d), and (f) are the spin Berry curvature for
ASHCs: $\sigma^y_{yx}$ of Mn$_{2}$Au (b), $\sigma^y_{yx}$ of Fe (d),
and $\sigma^y_{yx}$ of Pt (f). The black lines
denote the intersections of the Fermi surface with the slices.
The color code is on the log scale.}
\end{center}
\end{figure}

As discussed in the Supporting Information, the exchange-correlation
potential between spin-up and spin-down electrons in the Kohn-Sham equations depends on the magnetic order parameter (see Supporting Information).
Thus, the spin-polarized DFT calculations under the local spin
density approximation can be used to study order parameter effects on
exchange interactions and SOIs \cite{LSDA1, LSDA2}.
Hence, the calculated electronic structures, such as band structures
and wave functions, contain the effects of the order parameter.
Since the spin Berry curvature is directly calculated on the basis
of electronic structures [Eq. (5)], this shows an intrinsic
connection of the ASHE to the spin Berry curvature. In fact, the
intrinsic connection between the spin Berry curvature and
conventional SHE was also well justified in previous studies
\cite{Yao,Nagaosa,Niu}.

We further computed the $k$-resolved spin Berry curvatures for both
CSHCs (a, c, e in Fig. 2) and ASHCs (b, d, f in Fig. 2) of bulk
Mn$_{2}$Au, Fe, and Pt. A common feature is that the spin Berry
curvatures for ASHCs are much smaller than those for CSHCs, which
agrees with the experimental fact that the ASHCs in these materials
are smaller than the CSHCs. In addition, positive and negative spin Berry curvatures
coexist in the BZ, and both CSHC and ASHC values are determined by
the sum of all $k$ points. Thus, although the spin Berry curvature
value of Fe is larger than that of Mn$_{2}$Au throughout the BZ,
its ASHC is much smaller than that of Mn$_{2}$Au due to the
cancellation of positive and negative spin Berry curvatures.
This result also agrees well with the symmetry analysis above.
For Pt, the spin Berry curvatures for the CSHC are dominated by
regions with large positive values, while those for ASHC are
tiny with equal weights on the positive and negative values, leading
to the largest CSHC and a negligible ASHC in comparison with those
of Mn$_{2}$Au and Fe. The calculated Hall conductivities shown
in Table I are in good agreement with these spin Berry curvatures
profiles. The above results clearly show that the spin Berry
curvature is intrinsically connected not only to the conventional
SHE but also to the ASHE.

\begin{figure}[htb]
\begin{center}
\includegraphics[angle=0,width=0.95\linewidth]{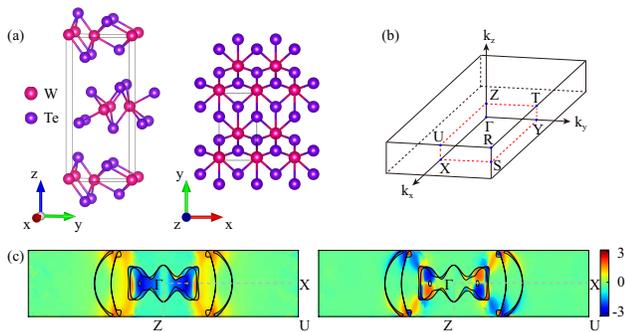}
\caption{\textbf{Atomic structure, BZ, and spin Berry curvature
of WTe$_2$.} Atomic structure (a), schematic diagram of the BZ (b),
and $k$-resolved spin Berry curvature on a log scale in a slice of
the 2D BZ at k$_y$ = 0 (C) for bulk WTe$_{2}$. The left and right
panels in (c) show the spin Berry curvature for $\sigma^z_{zx}$ and
$\sigma^z_{zy}$, respectively. The black lines show the intersections
of the Fermi surface with a slice of the BZ.}
\end{center}
\end{figure}

\section*{DISCUSSION AND CONCLUSION}

The ASHE is not limited to the magnetic order parameter;
any order parameter breaking the inversion symmetry can lead to
nonzero ASHC via the SOI (such effect in nonmagnetic materials is
denoted as the unconventional SHE). For example, the crystal structure of
nonmagnetic material T$_d$-WTe$_{2}$ [Fig. 3(a)] breaks the mirror
symmetry with respect to the xz plane and thus leads to broken inversion
symmetry. The broken symmetry in this case induces a nonzero conventional electric field in the $\vec y$ direction ($E_{eff}^y$).
According to the SOI Hamiltonian presented in Eq. (3), a sizable unconventional SHE spin current can still
be generated when the spin polarization direction is perpendicular
to $\vec y$ (see Supporting Information). To verify this, we performed DFT calculations
on the spin Berry curvature and ASHC of WTe$_{2}$ and found
$\sigma^{z}_{zx}$=18.99 $(\hbar/e)S/cm$, reaching 18\% of its
conventional SHC counterpart $\sigma^y_{zx}$ [103 ($\hbar/e$)S/cm]
\cite{Zhao11}. However, we found that $\sigma^z_{zy}$ has
a much smaller value [2.08 $(\hbar/e)S/cm$], which may be attributed
to the indirect interaction between $E_{eff}^y$ and $v_y$.
This result is consistent with experimental observations, where
field-free switching of perpendicular magnetization was observed
only with charge current in the $\vec x$ direction and not in the
$\vec y$ direction \cite{Ralph,Dash,Singh}. In addition, as shown
in Fig. 3, the positive spin Berry curvature for $\sigma^z_{zx}$ is
much larger than that for $\sigma^z_{zy}$. This result confirms the
intrinsic connection between the spin Berry curvature and
unconventional SHE in nonmagnetic materials.

In addition, we would like to discuss the novelty of this work
in comparison with previous studies \cite{Ebert,Tomohiro, Das,
Haney, Qu}. The notion of the ASHE was first proposed for spin current
generation in a ferromagnet from an AHE-induced charge current through
the conventional SHE (a phenomenon of SHE+AHE) \cite{Tomohiro, Das,
Haney}. However, the ASHE in our study resulted from the combined
effect of the order parameter and SOI. The physical origins of the two
theories are fundamentally different. Moreover, the two theories make
distinct predictions: The ASHE in Refs. \onlinecite{Tomohiro,Das,Haney}
predicts a nonzero spin current only for the noncollinear magnetic
order parameter and external electric field. In contrast, the ASHE
in this work additionally predicts that the collinear order parameter
and charge current generate a spin current with collinear propagation
direction and polarization. Unconventional components of the SHC tensor
were also discussed in Ref. \onlinecite{Ebert} from the symmetry
restriction point of view. Their analysis is valid only when the order
parameter is not involved in spin-charge conversion, in contradiction
to the ASHE in this work. In fact, the analysis of Ref. \onlinecite{Ebert}
fails to give rise to the ASHC component in Mn$_2$Au observed in
experiments \cite{Pan}. Note also that order parameter
dependence of the conventional SHE and AHE in ferromagnets was recently
observed in ferromagnets \cite{Qu}.

Before closing, we would like to emphasize that a dense
k-point mesh setting is necessary to guarantee the reliability of
the SHC and ASHC calculations.
For example, in Fig. S1 of the Supporting Information, we show the
ASHC $\sigma^y_{xy}$ of Fe calculated with
various k-point mesh settings. Clearly, $\sigma^y_{xy}$ is as large as
183.48 $(\hbar/e)S/cm$ for a small k-point mesh of $50\times 50\times 50$
while its converged value is approximately 1.0 $(\hbar/e)S/cm$, which can be
obtained when the k-point mesh is larger than $400\times 400\times 400$.
Therefore, the k-point mesh convergence test is crucial to the Hall
conductivity calculations. 

In summary, we have revealed the nature of the intrinsic ASHE:
The ASHE originates from the order-parameter-induced spin-dependent
electric field, which generates a spin current via the SOI.
The intrinsic relationship among the order parameter, spin Berry
curvature and ASHE is also revealed. The effects of the order parameter
on the electronic structure lead to a nonzero spin Berry curvature
and thus a sizeable ASHE. The order parameter that gives rise to the
ASHE is not limited to the magnetic order parameter. Any order parameter that can
lead to broken inversion symmetry should contribute to the ASHE.
This study is expected to provide an efficient way to search for
and optimize materials with large ASHE, which is particularly
important in the design of high-performance SOT spintronic devices.

We thank Prof. Jian Zhou and Dr. Yongliang Shi for valuable discussions.
This work is supported by the National Natural Science Foundation of
China (No. 12074301, 12004295, and 11974296) and Hong Kong RGC Grants
(No. 16300522, 16301619, and 16302321). P. Li thanks China's
Postdoctoral Science Foundation funded project (No. 2020M673364) and
the Open Project of the Key Laboratory of Computational Physical
Sciences (Ministry of Education). This research used the resources
of the HPCC platform in Xi'an Jiaotong University.

\section*{Supporting Information}
The supporting information is available online. The supporting
materials are published as submitted, without typesetting or editing.
The responsibility for scientific accuracy and content remains
entirely with the authors.

\end{document}